# Gestalt Driven Augmented Collimator Widget for Precise 5 DOF Dental Drill Tool Positioning in 3D Space

Mine Dastan*  Antonio Emmanuele Uva†  Michele Fiorentino‡

Polytechnic University of Bari

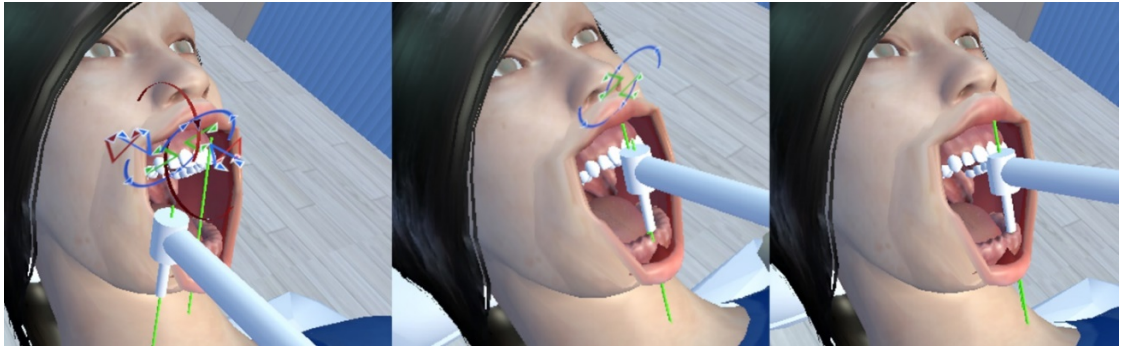

Figure 1: Our novel Augmented Collimator Widget (ACW) supports precise 5DOF drill tool positioning in dental implantology. ACW leverages the Gestalt principles, component separation and reification, error visual amplification, and dynamic hiding: approach (left), partial collimation (center), and full collimation, ACW disappears as not needed (right).

**ABSTRACT**

Drill tool positioning in dental implantology is a challenging task requiring 5DOF precision as the rotation around the tool axis is not influential. This work improves the quasi-static visual elements of the state-of-the-art with a novel Augmented Collimation Widget (ACW), an interactive tool of position and angle error visualization based on the gestalt reification, the human ability to group geometric elements. The user can seek in a quick, pre-attentive way the collimation of five (three positional and two rotational) error component widgets (ECWs), taking advantage of three key aspects: component separation and reification, error visual amplification, and dynamic hiding of the collimated components. We compared the ACW with the golden standard in a within-subjects (N=30) user test using 32 implant targets, measuring the time, error, and usability. ACW performed significantly better in positional (+19%) and angular (+47%) precision accuracy and with less mental demand (-6%) and frustration (-13%), but with an expected increase in task time (+59%) and physical demand (+64%). The interview indicated the ACW as the main preference and aesthetically more pleasant than GSW, candidating it as the new golden standard for implantology, but also for other applications where 5DOF positioning is key.

*e-mail: mine.dastan@poliba.it
† e-mail: antonio.uva@poliba.it
‡ e-mail: michele.fiorentino@poliba.it

## 1 INTRODUCTION

Minimal invasive implantology in dentistry increased in popularity thanks to its impact on patient health: reducing bone loss, surgical time, and tissue damage and improving aesthetics and psychological outcomes [1, 2]. Dentists must operate in a restricted oral cavity, and errors can lead to failure or even long-term damages. However, this medical procedure requires high spatial precision (see Fig. 2) with small margins of error in a very complex environment characterized by a non-fully sedated patient, soft tissues, limited working spaces, presence of sprays & liquids, and non-optimal lighting [3]. Due to the geometry of the drill tool, only five degrees of freedom (5DOF) must be guided as the rotation around the tool axis is not influential.

Errors in drill position or angle can lead to implant failure or even permanent damage (nerve, teeth, bone, and gingiva) and even continuous pain [4]. Robotic surgery technology is still not ready [5], and current implantology depends on the dexterity of the doctors. Current dental computer-assisted surgery (CAS) uses pre-operative CAD tools to identify the insert's locations from CT scans, and during the intervention, real-time guidance using 2D screens and 3D tracking [6]. Screen-based CAS demonstrated better performances [7], but the displays cannot be placed in the line of sight of the dentist, causing divided attention, hand & eye coordination adaptation, and long learning curves [8]. The continuous spatial mapping from 2D to 3D is cognitively demanding and can lead to distraction, fatigue, higher error rates, and psychological stress [9].

Augmented reality (AR) demonstrated its potential in assisting tool positioning in industrial (welding [10] and drilling[11]) and medical domains [12–14] by a visual overlay on the user's point of view [15]. However, AR approaches for drill positioning in literature use cognitive demanding numerical values [16] or more intuitive but quasi-static virtual 3D widgets to be aligned (see Fig. 4).

Our novel Augmented Collimation Widget (ACW) (see Fig. 1), takes inspiration from photographic and engineering instruments (e.g. viewfinder in analogic camera and the caliper), and leverages the following aspects: interactivity, analogic visualization, and the application of Gestalt principles. The "Gestalt" was coined in the 1920s by German psychologists [17] -means "unified configuration of forms or shapes"- it is a set of five principles, predicting humans' visual perception: proximity, similarity, continuity, closure, and connectedness. The human eye scans the scene and, according to Gestalt reification, the mind can identify specific patterns in spatial memory in less than 500 milliseconds, even adding details that are not in the original image (see Fig. 3).

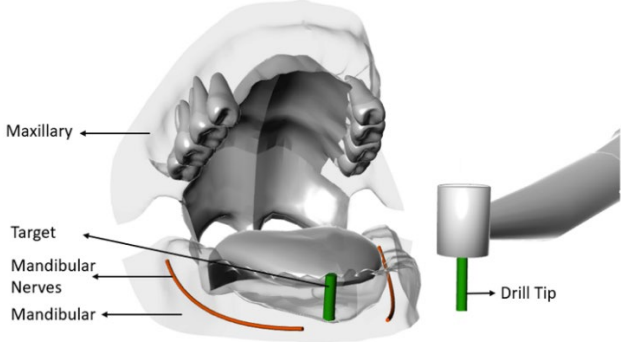

Figure 2: The main implantology problem: the drill must be positioned in the right position and angle in a narrow working environment and without damaging tissues, bones, or nerves.

In our ACW (detailed in section 3) the user pursues the analogic collimation of five couples of error components widgets (ECW) - three for positioning and two for rotation- in a pre-attentive way. The key innovative aspects of ACW are component separation and reification, error visual amplification, and the dynamic hiding of the collimated components. The user can seek in a quick, pre-attentive way the collimation of three couples of triangles for the spatial and two couples of semi-circles for the angular error, taking advantage of three aspects: component separation and reification, error visual amplification, and the dynamic hiding of the collimated components.

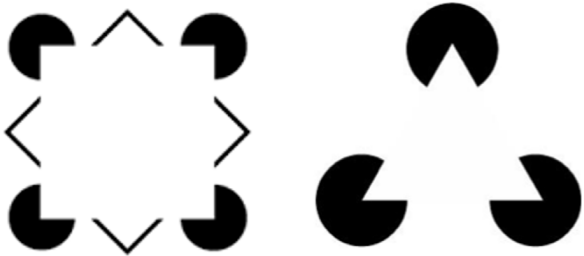

Figure 3: Gestalt Reification principle: the mind visualizes a white cube (left) and a triangle(right) in a quick, pre-attentive way.

In this paper, we present the ACW and compare it with the literature baseline for precision and accuracy, usability, and execution time. This work focuses only on visual design, without addressing tracking technology issues. It contributes with the following: i) identifies the golden standard of AR tool positioning in dental implantology from a systematic review, ii) presents the novel ACW design and implementation, iii) compares ACW to GSW with a user test.

## 2 STATE OF ART

We conducted a systematic review on the SCOPUS database (query: TITLE-ABS-KEY ("augmented reality" AND ("dent*" AND "implant*") AND NOT "Prosthetics")). From the initial 35 documents, we removed 4 related to maxillofacial surgery and 25 did not address tool guidance.

Ma L. et al. [18] supported dental drill placement by AR 3D registered visual aids of the "*Drill Tip*", "*Target*", "*Nerves*" and "*Mandible Model*". They compared AR with unsupported on five 3D printed models with ten parallel pins (simulate implants). AR-guided had better results (mean positional error = 1.25 mm vs. 1.63 mm; mean angle error = 4.03° vs. 6.10°).

Jiang W. et al. [19] improved AR drill placement utilizing the 3D visual assets "*Target*" and "*Drill Tip*" and dynamic color coding according to the user error (green-yellow-red) while "*Nerves*" are shown in static yellow. AR resulted in better performances (< 1.5-mm, < 5.5°).

Wang J. et al. [20] used AR to assist in drilling between two teeth roots. The virtual visual aids are "*Nerves*", "*Drill Tip*", "*Target Axis*" and "*Teeth Model*". The experiments lead to positive results (first experiment: 1.1 mm, 2°, second: 2.48 mm, angle not assisted).

Katic D. et al. [21] applied context-aware AR drill placement by "*Drill Tip*", "*Tool Axis*", "*Target*", "*Target Axis*" and "*Nerves*". The widget's color indicates the correct positioning, and a visual interface also suggests the upcoming drill head. Dentists tested it on a phantom model (positional error min 0.8, max 3.6mm, and angular min. 1.7, max. 6.5°) and a pig cadaver (1.1mm positional and 2.0 angular error). Dentists reported positive feedback on the simplicity of alignment, ergonomic, and cognitive effort.

Lin Y. et al. [22] presented AR drill placement with 3D visual assets, "*Drill Tip*", "*Target*", "*Target Axis*" and "*Nerves*". Six mandibular and four maxillary implants were inserted on a mouth model by a dentist (mandible error 0.50±0.33 mm, 2.70±1.55°, and maxillary 0.46±0.20 mm, 3.33±1.42°).

In conclusion, all the six AR applications in the literature -dated from 2010 to 2019- are summarized (Table 1), and reported better performance than the traditional (unassisted) methods (average positional error: 1.43 mm, angular 3.81°) [23].

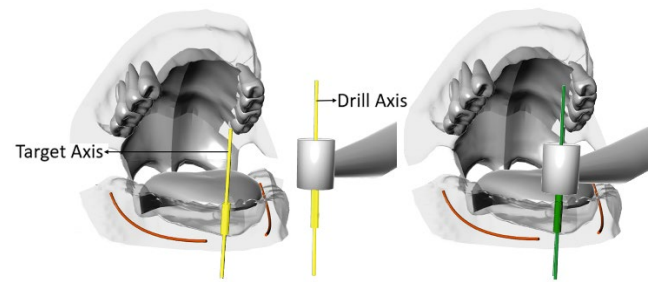

Figure 4: The golden standard widget (GSW): the user must align two cylinders attached respectively to the tool and the target.

From the visual interface design standpoint, we found a common approach we synthesized in the **Golden Standard Widget** (GSW). GSW consists of a pair of 3D widgets: a cylinder with an axis virtually attached to the drill tip along the rotational axis- and a similar one attached to the target calculated in the preoperative (see Fig. 4). The user must align the two widgets, supported also by dynamic coloring (red for misaligned and green to correct alignment). In our opinion, our novel ACW can improve GSW because it integrates different features in one single widget as detailed in the following section.

Table 1: State-of-the-art summary.

| Author | Target Mandible Model /subject | | Implant Number | AR Device | Frame Of Reference | Positional Error (mm) | Angular Error (°) |
|---|---|---|---|---|---|---|---|
| Ma L. et al. | 5x3D printed | 1 Human alive | 10 | IV Overlay Device | Screen Relative | 1.3 | 4.0 |
| Jiang W. et al. | 12x3D printed 1 phantom | 1 Human alive | 96 | 3D Image Display | Screen Relative | <1.5 | <5.5 |
| Wang J. et al. | 1 phantom | - | - | 3D Image Display | Screen Relative | Not available | |
| Katic D. et al. | - | 1 Pig cadaver | 2 | HMD | World Relative | 1.1 | 2.0 |
| Lin Y. et al. | 8x3D printed | - | 48 | Sony HMZ-T1 | World Relative | 0.5± 0.3 mand.  0.5± 0.2 maxillary | 2.7±1.6 mand. 3.3±1.5 maxillary |
| Katic D. et al. | 2 Phantom | - | 7 | Sony Glasstron | World Relative | Min. 0.8 Max. 3.6 | Min.1.7 Max 6.5 |

## 3 THE AUGMENTED COLLIMATOR WIDGET DESIGN

The design objective of the ACW is to visualize in a quick and pre-attentive way the positional (**pe**, vector) and angular error (**ae**, quaternion) of the tool compared to the designated target. ACW is virtually attached to the drill tool in the user's hand for leveraging proprioception [24]. In a preliminary exploration of the Gestalt theories, we quickly implemented and tested several concepts (see Fig. 5), and learning from mistakes, we refined the final design which integrates the three key aspects.

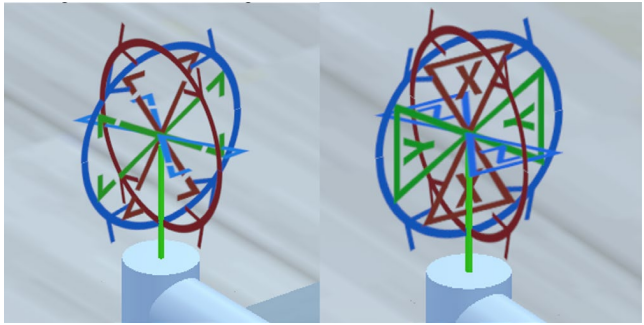

Figure 5: Two (of the many) discarded ACW prototypes; resulted in visually occluded, complex, or not being easy to use.

### 3.1 Error Components' Separation and Reification

ACW breaks down the 5DOF error visualization in spatial components. This approach is common in literature [25] and desktop 3D CAD systems, as it reduces cognitive and coordination effort and can improve precision and usability.

Positional (**pe**) and angular (**ae**) errors are defined as:

$$\mathbf{pe} = \mathbf{dp} - \mathbf{tp} \quad (1)$$

where **dp** and **tp** are the positional vectors in world space respectively of the drill tip and the implant target entry point designed from preoperative;

$$\mathbf{ae} = \mathbf{dq} * \mathbf{tq}^{-1} \quad (2)$$

**dq** and **tq** are the quaternions respectively of the drill tip and the implant target directions.

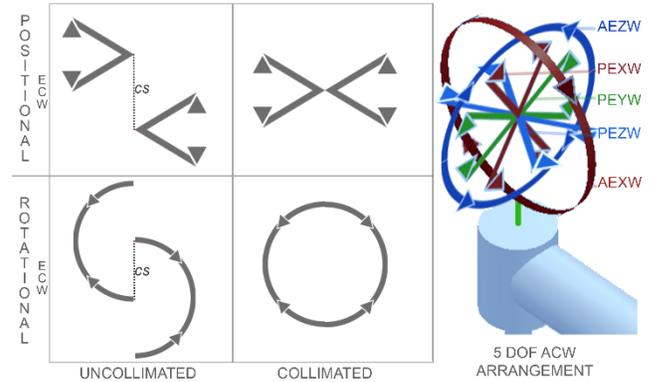

Figure 6: The error component widget (ECW) - before (left) and after (center) collimation- and a fully collimated ACW with all five ECWs assembled in 3D space (right).

We used five -conceptually similar- world referenced error component widgets (ECW): three for the positional, PEXW (horizontal, positive towards the right side of the user initial position), PEYW: (positive up), PEZW (positive front), and two for angular AEXW (pitch), and AEZW (roll).

Each ECW is quickly identified by a pair of facing symbols laid on the same plane and moving symmetrically along the component direction or axis: ">" for positional and "C" for angular error and color (red, green, and blue). The ECW symbols are chosen to convey strong reification when collimated along a component: position collimation forms an equally colored "X", and angular a full circle. The ECW collimation indicated a perfect positioning of that component. After initial tests, we added some small arrows to the original symbols to have a quick hint of the direction associated with the specific ECW.

The five ECWs (all together compose the ACW) are arranged in the space according to their spatial meaning and scaled to be all visible before and during the collimation (Fig. 6).

### 3.2 Error Visual Amplification

This aspect mimics common dentists' practice of using zoom lenses. We used visual amplification of the ECW error response to overcome the physical resolution of the AR display and limited human capability to perceive the error. We define the Collimator separation ($cs_i$) as the linear distance in the virtual world space of each ECW (i) symbols pairs (see Fig. 6).

We implemented a piecewise function (3) to amplify the visual for each $ECW_i$ where $Gain_i$ is the linear amplification factor, and $e_i$ is the scalar value of the specific $ECW_i$ (distance or Euler angle). A positive gain will amplify the error visualization as seen by the user during the collimation.

$$cs_i = \begin{cases} 0 & \text{if} \quad e_i \leq acce_i \\ Gain_i \times e_i & \text{if} \quad ae_i < e_i < mdt_i \\ Gain_i \times mdt_i & \text{if} \quad e_i \geq mdt_i \end{cases} \quad (3)$$

In early tests, we observed that when the drill is far away from the target (e.g., approach phase), the $cs$ was large and the widget invisible because the two halves are out of the user's sight frustum. Therefore, we implemented for each ECW a $mdt_i$, a max distance threshold, that clamps the amplification at larger error values, where the visual approach is sufficient and even more effective than ACW.

Similarly, we clamp the amplification to 0 when the $acce_i$, the accepted error is reached, to avoid the user's struggle to seek an unnecessary and impossible-to-reach ideal value (see Fig. 7).

### 3.3 Dynamic hiding of the collimated components

Since early designs, we observed that in near collimation conditions, the ACW widget becomes visually crowded and complex to understand. To address this issue, we firstly located ACW on the top of the tool for better visibility of the working area, a safety measure not considered in previous works.

In addition, we reduced the visual and cognitive effort of the user, by hiding the ECWs that are below the accepted error $acce_i$. In practice, the user, after initial training will use his pre-attentive skills to make all the ECW collimate and disappear (see Fig. 8).

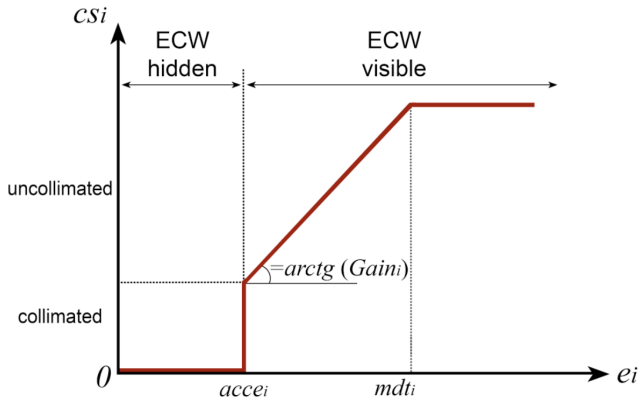

Figure 7: ECW visual amplification behavior ($cs_i$) with the error ($e_i$).

### 3.4 ACW Implementation

We implemented the ACW with a modular approach, designing a unique Unity 3D ECW prefab. At runtime, when the user grabs the drill and moves it in space, the main script triggers an event called tool "position modified" which includes the positional and angular error components compared to the current target position.

Each ACW's five ECW prefab is instantiated with a different texture, spatial orientation, and behavior parameters ($Gain_i$, $acce_i$, $mdt_i$) and is connected to the related components. This approach allowed us in the initial phase to try and evaluate in a short time different visual configurations and behaviors in a heuristic way, to find the optimal final design used in the experiment.

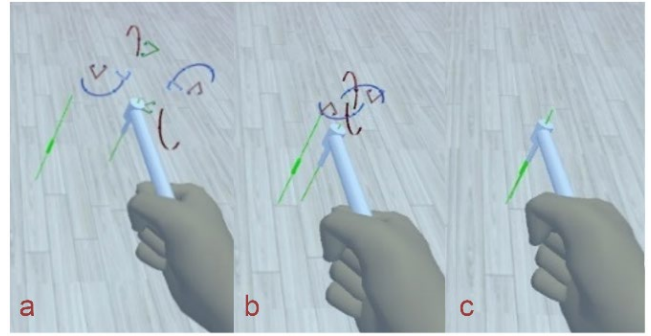

Figure 8: ACW dynamic hiding: all 5 ECWs are uncollimated and visible (a), PEYW and PEZW disappear (b), and the target is fully acquired (c).

## 4 EXPERIMENT DESIGN

We conducted a within-subjects experiment to compare ACW to GSW. We choose to simulate AR simulation in VR because of the Covid19 pandemic limitations and to obtain a consistent and repeatable setup as reported in the literature [26]. The study was compiled with the American Psychological Association Code of Ethics and informed consent was obtained from each participant.

We set five hypotheses:

**H1.** ACW improves positional precision and accuracy vs GSW;
**H2.** ACW improves angular precision and accuracy vs GSW;
**H3.** ACW reduces execution time vs GSW;
**H4.** ACW reduces the mental load vs GSW;
**H5.** ACW improves perceived performance vs GSW.

### 4.1 Participants

30 unpaid voluntary subjects were recruited from a local university from 21 engineering and 9 medicine students (33.3% male and 66.7% Female), aged between 21 to 34 Years (27 average and 3,188 standard deviations), 90% are right-handed and 10% are left-handed, and 3 participants develop AR/VR applications, 15 novice AR/VR users.

### 4.2 Setup

Users sit on a stool like a dentist's practice in front of a table (see Fig. 9). The VR system is the Oculus Quest 2 and dual touch controllers. The dominant hand controller grip button is used for grabbing the drill tool and the trigger button for confirming the target position. The non-dominant controller has a ray cast that uses the trigger button to interact with the 3D menus.

We developed a standalone application on the Oculus using Unity 3D to perform the test procedure (see Fig. 10), with a welcome room with visual instructions and 3D buttons to select one of the two treatments. We created a training scene composed of a 4mx3m room with a table and 32 random targets in a sphere of a 30 cm radius centered on the initial tool position. The test scene is composed of an identic room but instead of the table, it shows a static patient avatar with a mouth open and 32 target sets (one for each tooth, 16 maxilla, and 16 mandibular) positioned by an experienced dentist. The application in both scenes displays all the targets randomly and once per time. Graphical elements transparency was set to be visually equivalent to a HoloLens 2 display. When the user is confident of the positioning, the user hits the trigger button, and the app saves the captured metrics in a comma-separated file.

We set the following ACW parameters:
$acce_i$ = 2mm, $mdt_i$ = 50mm, $Gain$ = 50 for positional ECWs,
$acce_i$ = 2°, $mdt_i$ = 45°, $Gain$ = 0.1 for angular ECWs

Table 2: Positional error descriptive statistic.

|  | pem | | $pe_x$ | | $pe_y$ | | $pe_z$ | |
|---|---|---|---|---|---|---|---|---|
| Widget Type | ACW | GSW | ACW | GSW | ACW | GSW | ACW | GSW |
| Valid | 900 | 900 | 900 | 900 | 900 | 900 | 900 | 900 |
| Mean | 2.24 | 2.73 | 0.60 | -0.43 | -0.53 | -0.45 | -0.08 | -0.48 |
| Std. Deviation | 1.42 | 2.75 | 1.44 | 2.57 | 1.64 | 2.41 | 1.28 | 1.41 |
| Skewness | 3.25 | 16.57 | 0.88 | -13.29 | -0.07 | -2.73 | 1.41 | 2.02 |
| Std. Error Skewness | 0.08 | 0.08 | 0.08 | 0.08 | 0.08 | 0.08 | 0.08 | 0.08 |
| Kurtosis | 18.69 | 389.95 | 7.68 | 313.10 | 5.11 | 35.71 | 15.32 | 18.84 |
| Std. Error of Kurtosis | 0.16 | 0.16 | 0.16 | 0.16 | 0.16 | 0.16 | 0.16 | 0.16 |
| Shapiro-Wilk | 0.76 | 0.38 | 0.94 | 0.48 | 0.94 | 0.87 | 0.91 | 0.90 |
| P-value of Shapiro-Wilk | < .001 | < .001 | < .001 | < .001 | < .001 | < .001 | < .001 | < .001 |
| Minimum | 0.17 | 0.17 | -4.88 | -59.52 | -9.99 | -32.64 | -8.03 | -4.71 |
| Maximum | 15.33 | 69.60 | 11.58 | 17.76 | 8.60 | 9.08 | 13.35 | 15.38 |

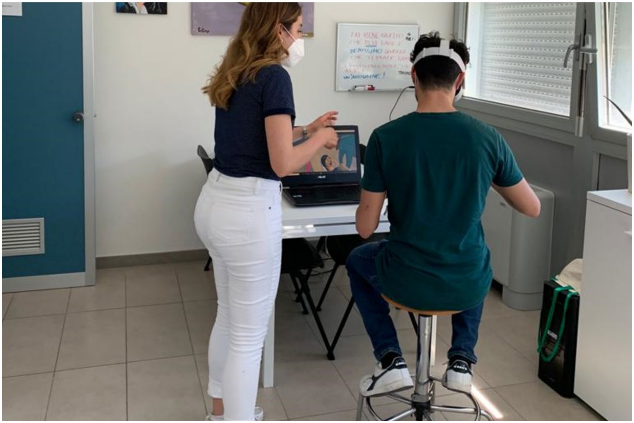

Figure 9: The experiment setup: the participant sits on a stool like a dentist, and wears Oculus Quest 2. An assistant supervises the experiment.

### 4.3 Procedure

Each subject receives a preliminary introduction about the test objectives and its procedures, gives its consent, and declares the absence of any general discomfort, fatigue, eye strain, difficulty focusing, or headache before the experiment.

The participant watches an introductive video that explains how the two widgets work and the interface of the experiment application. Subsequently, the participant is asked to sit on the footstool used by dentists, wear the Oculus Quest 2, familiarize themselves with the virtual reality environment and controls, and choose a designated treatment set to balance the conditions, according to a Latin square order:
- A: GSW training, ACW training, Mandibular (MD) GSW, Maxillary (MX) GSW, Mandibular (MD) ACW Maxillary (MX) ACW;
- B: ACW training, GSW training, MD-ACW, MD-GSW, MX-ACW, MX-GSW.

During the training session, the targets are located in the space while in the other sessions the targets' location changes in mandible or maxillary.

Participants were not allowed to use the desk as arm support, to imitate the dentist's real conditions, and were asked to focus on the maximum possible precision of the drill instead of the task time. When one treatment is finished the application loads automatically the next one in the set. Due to the short duration of the experiment, there are no programmed pauses, however, participants are informed that they can stop the experiment at any time. An assistant was always available and supervised all the operations, taking notes of unexpected situations.

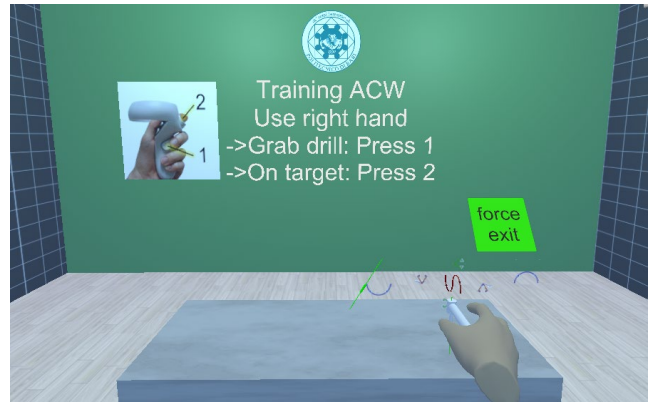

Figure 10: The experiment welcome room: instructions are displayed on the front wall and the user interacts with the drill tool using the dominant-hand controller and non-dominant to interact with the 3D menu buttons.

After completing the task, the participant takes off the headset and is asked to compile an online questionnaire. We implemented all health and hygiene protocols using disposable masks and sanitizing both devices and places.

### 4.4 Metrics

We acquired the following metrics using the test application:
**Positional Error magnitude (pem)** scalar value in mm,
**Positional Error component ($pe_{x|y|z}$): pe** cartesian world referenced component,
**Angular Error magnitude (aem): ae** rotation angle in degrees,
**Angular Error components ($ae_{x|y|z}$): ae** rotation angle component in degrees,
**Task Time (tt):** time elapsed between the new target display and user button press in milliseconds.

For qualitative evaluation of the user experience, we acquired:
**NASA TLX (Mental, Physical Temporal Demand)** [27];
**Perceived Performance Rate (Likert 7 scale);**
**Aesthetical Rate (Likert 7 scale);**
**Preferred widget (Likert 7 scale).**

Table 3: Angular error descriptive statistics.

|  | aem | | $ae_x$ | | $ae_y$ | | $ae_z$ | |
|---|---|---|---|---|---|---|---|---|
| Widget Type | ACW | GSW | ACW | GSW | ACW | GSW | ACW | GSW |
| Valid | 900 | 900 | 900 | 900 | 900 | 900 | 900 | 900 |
| Mean | 5.03 | 9.55 | 0.43 | 1.54 | -3.80 | -4.69 | -0.72 | -3.42 |
| Std. Deviation | 3.14 | 5.77 | 1.28 | 4.23 | 4.03 | 6.46 | 1.46 | 5.38 |
| Skewness | 1.30 | 1.91 | -0.80 | 0.03 | 0.05 | -0.34 | -1.08 | -0.45 |
| Std. Error Skewness | 0.08 | 0.08 | 0.08 | 0.08 | 0.08 | 0.08 | 0.08 | 0.08 |
| Kurtosis | 2.09 | 6.01 | 8.27 | -0.02 | 2.36 | 3.51 | 6.37 | 2.07 |
| Std. Error of Kurtosis | 0.16 | 0.16 | 0.16 | 0.16 | 0.16 | 0.16 | 0.16 | 0.16 |
| Minimum | 0.33 | 0.52 | -11.03 | -11.99 | -22.07 | -35.16 | -11.03 | -28.87 |
| Maximum | 22.07 | 39.31 | 7.08 | 14.75 | 17.155 | 18.133 | 4.362 | 15.569 |

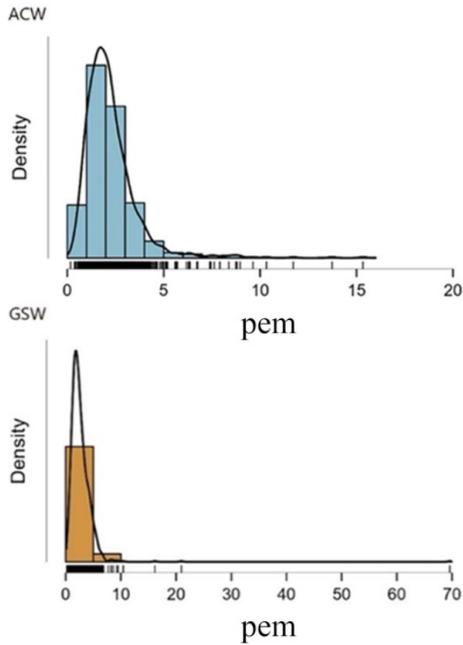

Figure 11: Distribution Plots for positional error magnitude (pem).

## 4.5 Results

All (N=30) participants performed the training and the experiment sessions without interruptions or anomalies. The test (excluding the training) average time was 6.10±4.13 minutes.

We analyzed the total 1920 (30*2*32) observations for outliers and removed from each participant data all the first target of each treatment (30*2*2) as the assistant notes and statistical analysis found significant differences both in time and precision compared to subsequent targets. This anomaly can be explained by the user interacting with the "next treatment" button or the drill-grab required at every treatment change. The remaining 1800 entries (30*2*30) were analyzed.

### 4.5.1 Positional Error

The descriptive statistics (Table 2) show similar skewed behavior for both treatments, supported by the rejection of Shapiro-Wilk's test of normality and the Equality of Two Variances (Levene's). Therefore, we apply the nonparametric independent Mann-Whitney U one tail T-test, indicating ACW is significantly more precise and accurate (2.24±1.42 mm for ACW and 2.72±2.75 mm for GSW) than GSW (W = 331360.000, p< .001) (see Fig. 11). Studying the positional error components ($pe_{x|y|z}$) we also can see that all are rejecting the null hypothesis for normality, and Mann-Whitney U one tail T-test finds that the $pe_x$ and $pe_z$ differ in the two groups.

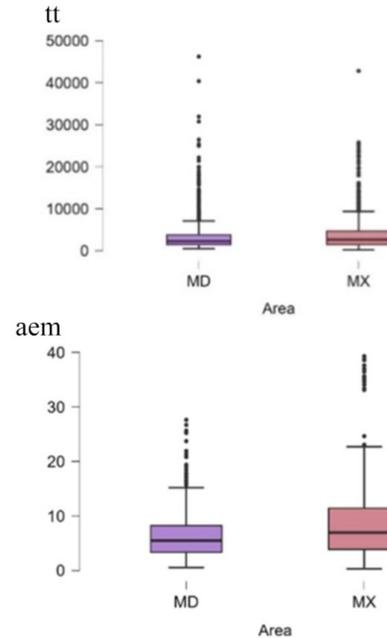

Figure 12: Box Plots of tt and aem for Mandibular (MD) and Maxillary (MX)

Table 4: Task Time (tt) descriptive table for ACW and GSW

|  | tt | |
|---|---|---|
| Widget Type | ACW | GSW |
| Valid | 900 | 900 |
| Mean | 5194.40 | 2141.88 |
| Std. Deviation | 4918.58 | 1817.63 |
| Skewness | 3.36 | 2.77 |
| Std. Error Skewness | 0.08 | 0.08 |
| Kurtosis | 16.07 | 9.33 |
| Std. Error of Kurtosis | 0.16 | 0.16 |
| Minimum | 203.00 | 278.00 |
| Maximum | 46219.00 | 14074.00 |

### 4.5.2 Angular Error

The descriptive statistics (Table 3) showed similar skewed behavior in both groups, with a sample mean of 5.03±3.13° for ACW and 9.54±5.77° for GSW. Shapiro-Wilk's and Levine's tests are not rejected, indicating that the group data might not be normally distributed, and variances are unequal. Therefore, we apply a nonparametric independent one-tail T-test Mann-Whitney U that indicating ACW is significantly more precise and accurate than GSW (W=178363.500, p< .001) (see Fig. 13). The box plot results also showed that the mandibular condition is more precise than the maxillary (see Fig. 12).

Table 5: Descriptive statistics of NASA-TLX post-experiment questionnaire.

|  | Mental D. | | Physical D. | | Temporal D. | | Performance | | Effort | | Frustration | | Aesthetical | |
| --- | --- | --- | --- | --- | --- | --- | --- | --- | --- | --- | --- | --- | --- | --- |
| Widget | ACW | GSW | ACW | GSW | ACW | GSW | ACW | GSW | ACW | GSW | ACW | GSW | ACW | GSW |
| Valid | 30 | 30 | 30 | 30 | 30 | 30 | 30 | 30 | 30 | 30 | 30 | 30 | 30 | 30 |
| Mean | 2.50 | 2.67 | 4.06 | 2.47 | 2.90 | 2.43 | 4.60 | 4.70 | 4.17 | 2.73 | 2.00 | 2.03 | 4.83 | 5.13 |
| Std. Deviation | 1.49 | 1.01 | 1.37 | 1.28 | 1.45 | 1.25 | 1.25 | 1.32 | 1.05 | 1.26 | 0.83 | 0.97 | 1.09 | 1.17 |
| Minimum | 1 | 1 | 1 | 1 | 1 | 1 | 1 | 1 | 3 | 1 | 1 | 1 | 2 | 3 |
| Maximum | 6 | 5 | 7 | 5 | 6 | 5 | 7 | 7 | 7 | 5 | 4 | 4 | 7 | 7 |

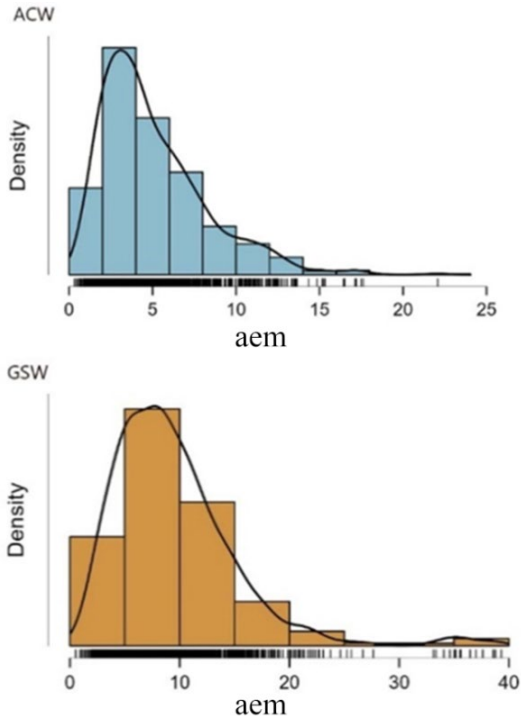

Figure 13: Distribution Plots for Angular Error magnitude (aem)

Studying the angular error components ($ae_{x/y/z}$) we also can see that all are rejecting the null hypothesis for normality, and Mann-Whitney U one tail T-test finds that the $ae_x$ and $ae_z$ significantly differ in the two groups (see Fig. 14).

### 4.5.3 Task Time

The descriptive statistics (Table 4) showed similar skewed behavior in both treatments, with a sample mean of 5194.4±4918.579 milliseconds for ACW and 2141.882±1817.623 milliseconds for GSW. Shapiro-Wilk's and Levene's test yields are not rejected, indicating that the group data might not be normally distributed, and variances are unequal. Therefore, we apply the nonparametric independent one-tail t-test Mann-Whitney U indicating ACW is significantly slower than GSW (W = 679069.000, p< .001). The box plot results showed that the mandibular condition is more precise than the maxillary (see Fig. 12).

### 4.5.4 NASA TLX

According to the answers (Table 5), the ACW resulted to have less mental demand (Likert 2.50/7 vs 2.67/7, -6%) and frustration (Likert scale 2.00/7 vs 2.30/7, -13%) in the participant compared to GSW. This was an unexpected result, considering that the widget is more complex. We can explain this outcome by the fact that this interface. is more supportive and intuitive. We also can see that all are rejecting the null hypothesis for normality, and Mann-Whitney U one tail T-test finds that the Physical demand (W=715.00 p< .001) and effort differ in the two groups (W=709.00 p< .001).

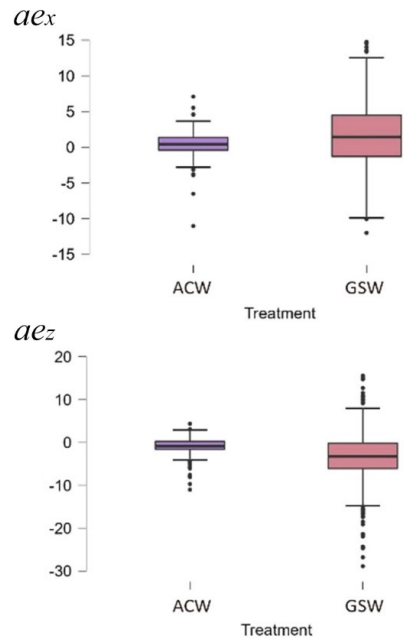

Figure 14: Box plots for $ae_x$ and $ae_z$ components.

ACW required more physical (Likert scale 4.06/7 vs 2.47/7, +64%), temporal demand (Likert scale 2.90/7 vs 2.43/7, +19%), and effort (Likert scale 4.17/7 vs 2.73/7, +53%) (see Fig. 15). This result was expected and confirmed by the quantitative measurements indicating the longer time of positioning with unsupported limbs in the air.

Participants barely perceive their superior performance using ACW (Likert scale 4.7/7 vs 4.6/7, +2%). The users claimed that the ACW required more time and that ACW is more sensible for wrist movement, probably meaning helping better to achieve angular positioning. The interviews are coherent with the real performance, a nontrivial as in many other VR applications the user perception differs [28].

### 4.5.5 Preferred widget

Most of the participants prefer the ACW (19/30, 63.3%) and its aesthetics (Likert 5.13/7 vs 4.83/7, +6%), and one user claimed that ACW does not require extra body movement and the task was easier compared with GSW and its thin axes. Many participants reported that ACW was tough to learn initially but over time they adapted and feel faster during the experiment.

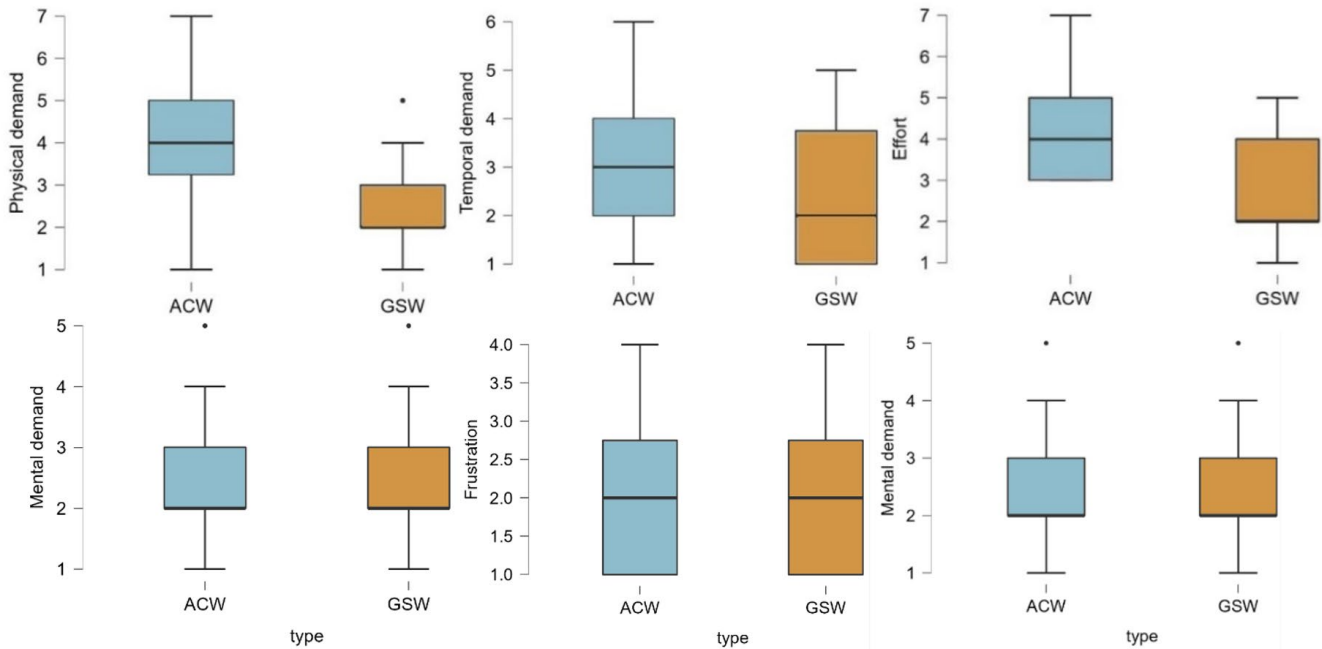

Figure 15: Box plots results for NASA-TLX.

## 5 Discussion

The results support the first two hypotheses: ACW provides better 5DOF positional and angular precision and accuracy (H1&H2), with an advantage for roll and pitch angle and along X and Y components. The angular error is key in the implant placement because it can lead to wrong tool paths during drilling, secondly, the widget can help in the complex environment where visibility can be reduced.

The hypothesis claiming ACW is faster (H3) is not supported. This aspect is partially expected as the user must pay much more attention, dexterity, and time. However, in the specific implantology task, precision is the most important aspect, and physical and time demands are secondary. The two hypotheses ACW reduced the mental load (H4), and performance improvement of ACW compared to GSW (H5) are also supported.

We can conclude how the experiment outcomes, even with similar but not equal modalities and scenarios, are in good agreement with the advantage provided by AR interfaces reported in the literature [18, 19]. The questionnaire evidenced less mental demand and frustration for the ACW, and the participants prefer ACW for performance and aesthetics, indicating the validity of the design and its principles.

## 6 Conclusion

We presented a novel collimator widget (ACW), an interactive 5DOF position and angle error Augmented reality (AR) visualization tool for dental implantology applications. ACW is based on the gestalt reification, the human ability to group geometric elements. The user can seek in a quick, pre-attentive way the collimation of five (three positional and two rotational) error component widgets (ECWs), taking advantage of three key aspects: component separation and reification, error visual amplification, and dynamic hiding of the collimated components. Initially, we identified the golden standard widget (GSW) for dental tool positioning in the literature. Then we presented the design principles and methodology of ACW. We performed a within-subjects (N=30) user test using 32 implant targets, measuring the time, error, and usability. ACW performed significantly better, especially in angular (+47%), but also in positional (+19%) precision accuracy and with less frustration (-13%) and mental demand (-6%), but with an expected increase physical demand (+64%) and in task time (+59%). The interview indicated the ACW as the main preference and aesthetically more pleasant than GSW. This work has also revealed the importance of the angular error, its implication, and, how the GSW can be improved mostly in this aspect.

We think that this work can have still margins of improvement. Our widget implementation and the experimental infrastructure built in Unity 3D are designed to be very flexible, and this allows us to test rapidly different configurations. In future works, we want to test ACW in an AR setup and with dentists and explore the different configurations of gain functions (e.g. nonlinear) and parameters.

In conclusion, we are very encouraged by the positive outcomes of this study and believe that the ACW can become the new golden standard not only for implantology but also it can be extended to 6DOF and interest a wider number of applications in medical and industrial fields (CAD, welding, assembly, etc.).